\documentstyle[12pt,aasms4]{article}

\def\reference{\par\noindent\hangindent=1cm\hangafter=1}

\newcommand{\eq}{\begin{equation}}
\newcommand{\ee}{\end{equation}}

\def\t0{\theta_{\circ}}

\def\be{\begin{equation}}
\def\en{\end{equation}}

\def\gapp{\ \lower 3pt\hbox{${\buildrel > \over \sim}$}\ }
\def\lapp{\ \lower 3pt\hbox{${\buildrel < \over \sim}$}\ }
\newcommand{\pa}{\partial}
\newcommand{\beq}{\begin{equation}}
\newcommand{\eeq}{\end{equation}}
\lefthead{Yeh \& Jiang}
\righthead{Scattered Planets}

\begin{document}

\title{Orbital Evolution of Scattered Planets}

\author{Li-Chin Yeh \& Ing-Guey Jiang}

\affil{Institute of Astronomy and Astrophysics, 
Academia Sinica, Taipei, 
Taiwan}

\authoremail{jiang@asiaa.sinica.edu.tw}

\begin{abstract}

A simple dynamical model is employed to study the possible orbital 
evolution of scattered
planets and phase plane analysis is used to classify
the parameter space and solutions. 
Our results reconfirm that there is always an 
increase in eccentricity when the planet was scattered to 
migrate outward when the 
initial eccentricity is zero. Applying our study on the Solar System
and considering the existence of the Kuiper Belt, 
this conclusion implies 
that Neptune was dynamically coupled
with the Kuiper Belt 
in the early phase of the Solar System, which is consistent 
with the simulational model in Thommes, Duncan \& Levison (1999).  
 
\end{abstract}


\section{Introduction}

Due to the observational effort, both the number of extra-solar planets 
and small bodies in the outer Solar System
has increased dramatically. These new discoveries
provide interesting laboratories to study theories of planetary 
dynamics (Jiang \& Ip 2001). 
On the other hand, some of the observational phenomena require   
explanation from the dynamical models. For example, 
in order to explain the dynamics of Plutinos,  
the possible migration of Neptune's orbit has been suggested 
by Malhotra (1995). Fernandez \& Ip (1984) proposed that the interaction 
with planetesimals might make the orbit of Neptune expand and therefore
migrate slowly. This ``slow migration'' was demonstrated by 
Hahn \& Malhotra (1999).

On the other hand, 
Thommes, Duncan \& Levison (1999) successfully 
showed that the Neptune might form in
the Jupiter-Saturn region of the Solar System and was scattered
to migrate outward to the current location due to the interaction
with the growing Jupiter or Saturn. In their model, Neptune was scattered
to have a ``fast migration'' at first and then transfer to 
``slow migration'' due to the interaction with planetesimals.

Numerical simulations provide opportunities for us to explore
the possible solutions for astronomical problems and are always 
good tools for more practical modelling. Nevertheless, it is important
to use the analytical model to classify the solutions on the parameter
space and to understand the principle
physical processes within the simulations.

Considering two planets moving around a central star, 
if the apocentre-distance of the inner planet is much smaller than
the pericentre-distance of the outer planet, there would be no scattering 
between these two planets. They can only influence each other through 
secular perturbation. Therefore, 
the scattering might happen when the inner planet (say, the smaller one)
moving on a more eccentric orbit and the outer planet (say, the giant planet)
moving on a circular orbit and their orbits cross each other. 
We can call this process 
``crossing scattering'' and the Tisserand relation 
is valid if the distance between these two planets
never gets too close. 

Another type of scattering process might happen for two planets 
both moving on circular orbits, as the newly formed Neptune and Saturn 
in Thommes, Duncan \& Levison (1999). Let us assume the small planet's 
orbital radius ($r_s$) is slightly shorter than the giant planet's
orbital radius ($r_g$) and so the angular velocity of the small planet
($d\theta_s/dt$) is slightly larger than 
the giant planet's angular velocity ($d\theta_g/dt$). If both of them moving 
anti-clockwise as the convention of a polar coordinate, there must be 
a chance that they approach to each other and go through three periods: 
(1) $\theta_s = \theta_g-\epsilon$, (2) $\theta_s = \theta_g$ and 
(3) $\theta_s = \theta_g+\epsilon$, where $\epsilon$ is a small number. 
During the first period, the small planet 
should gain the angular momentum from the giant planet because the giant planet
is pulling the small planet in the direction which the small planet is moving 
to. Similarly,  
during the final period, the small planet 
should lose the angular momentum to the giant planet because the giant planet
is pulling the small planet in the direction oppositive to
the small planet's moving direction. Obviously, during the middle
period, there should be no angular momentum exchange and
the interaction between two 
planets is stronger because they are closer to each other.
The net effect of this type of scattering is that the small planet is pulling
away from the central star and the angular momentum is approximately 
conserved. We can call this process 
``swing scattering''. 

If one assume that all planets are moving on circular orbits initially as in
Thommes, Duncan \& Levison (1999), this ``swing scattering'' definitely 
happen before the ordinary ``crossing  scattering'' could take place.
Therefore, both of them are important for the overall orbital evolution.

In this paper, we will focus on ``swing scattering'' only. 
We analytically model this process by adding 
a forcing term on the gravitational force
from the central star. This term represents 
the interaction with the growing giant planets.
We assume that the scattering is in the radial direction, so the angular 
momentum is still conserved for the planet. 



 We study 
the possible outcome under this scattering. Thus, in our system, 
the planet would face three stages: (1) pre-scattering stage: 
freely orbiting a central star in a two-body dynamical system. 
(2) scattering stage: influenced by the scattering
term in additional to the force from the  central star. 
(3) post-scattering stage: the scattering
stops and return to a two-body system again.

In Section 2, we describe the equations for our model. 
We do the phase plane analysis in Section 3 and report 
the results in Section 4. In Section 5, 
we provide the conclusions.

\section{The Model}

%
%

We consider a two-body system that a planet is moving around a central star
on the two dimensional space
governed by the attractive inverse square law of force with an additional
term representing the scattering. Therefore,
\beq
 f(r)=\frac{-k}{r^2}+k_1,\label{eq:force}
\eeq
where $k$ is positive and the minus sign ensures that the force is toward
the central star. Moreover, the parameter
$k_1$ is also positive, representing the 
scattering force in the direction 
away from the star. We assume that $k_1$ is a constant during the scattering.

Let $u=1/r$, from Goldstein (1980), we have the following equation
\beq
\frac{d^2u}{d^2 \theta}=-u-\frac{mf\left(1/u\right)}{l^2u^2},
\label{eq:1}
\eeq
where $m$ is the mass and $l$ is the angular momentum of the planet. 
We use the polar coordinate $(r,\theta)$ to describe the 
location of the planet.

The angular momentum $l$ is a constant here, so we have 
\beq
mr^2{d\theta}=l{dt}. 
\eeq
Because of this, we can use $\theta$ as our independent variable.
We use $\theta$ to label time $t$ afterward and one can easily
gets $t$ from the above equation.
 
From  Equation (\ref{eq:force}) and Equation (\ref{eq:1}), we have
\beq
\frac{d^2 u}{d^2 \theta}= -u+\beta-\frac{\beta_1}{u^2},\label{eq:2}
\eeq
where $\beta=m k/{(l^2)}$ and $\beta_1=m k_1/{(l^2)}$. One can see 
that $u$, $\beta$, and $\beta_1$ are all positive. 

We then transform Equation (\ref{eq:2}) to the following autonomous
system:
\begin{eqnarray} 
 \frac{du}{d\theta} &=&v \equiv f_1 (u,v)\nonumber\\
& &  \label{eq:3}\\
\frac{dv}{d\theta}& =& -u+\beta-\frac{\beta_1}{u^2}\equiv f_2(u,v). \nonumber
\end{eqnarray}

In order to understand how
the solutions depend on the parameters, we will do the phase plane
analysis to classify the solutions completely 
in terms of $\beta$ and $\beta_1$.

During the pre-scattering and post-scattering stage
when $\beta_1=0$, 
there is one fixed point $(u,v)=(\beta,0)$ on the phase plane. 
The eigenvalues corresponding to this fixed point are complex, 
i.e. $\lambda=\pm i$,
so it is a stable center point. In the next section, we begin to 
do the analysis for the scattering stage when $\beta_1 > 0$.

\section{Phase Plane Analysis}

The fixed points $(u,v)$ of Equation (\ref{eq:3}) satisfy the following 
equations:
\begin{eqnarray}
 & &v=0,   \nonumber     \\
 & &u^3-\beta u^2+\beta_1=0,\label{eq:4}
\end{eqnarray}
where $u\ne 0$. Let 
\beq
 y=-\frac{3u-\beta}{2\beta},\label{eq:y_tran} 
\eeq
equation (\ref{eq:4}) becomes 
\beq
4y^3-3y=\Delta, \label{eq:y}
\eeq
where 
\beq
\Delta=\frac{27\beta_1-2\beta^3}{2\beta^3}.
\eeq


The solutions of the above cubic equation completely depend on $\Delta$
(Neumark 1985). 
 We will consider all different values of $\Delta$  
except there are a few cases which are not inside our scope.
We will not consider the case when $\Delta<-1$ because this gives us
 $\beta_1<0$ but we assume  $\beta_1>0$ in this paper.
On the other hand,  
the probability that $\Delta=1$ and $\Delta=0$ are very small because 
the scattering parameter $\beta_1$ need to precisely 
equal to  particular number for a given $\beta$. 
We will not discuss the case of $\Delta=1$
because the eigenvalues are zero for the first order linearization
of the phase plane analysis.
The results for  $\Delta=0$ will be in Appendix A.

In fact, the value of $\Delta$ has the physical meaning of the 
strength of the scattering relative to the strength of the central 
star's gravity. Thus, $\Delta=-1$ means there is no scattering and
the larger $\Delta$ corresponds to the 
stronger scattering relatively to
the gravity of the central star.  

 Therefore, in this section, 
 there would be three cases, which correspond to different relative strength
of the scattering
: $\Delta>1$ 
($27 \beta_1-4\beta^3 >0$),
$0< \Delta<1$ ($2\beta^3<27 \beta_1<4\beta^3 $) and $-1< \Delta<0$
($0<27 \beta_1<2\beta^3$). 

\subsection{Case 1: $\Delta >1$} 

In this case, the solution of equation (\ref{eq:y})
is
\beq
y=\cosh \eta,
\eeq
where $\eta$ satisfies 
\beq
\cosh 3\eta=\Delta=\frac{27\beta_1-2\beta^3}{2\beta^3}. 
\eeq
Equation (\ref{eq:y_tran}) gives
\beq
u=\frac{\beta}{3}(-2y+1).\label{eq:u}
\eeq
Therefore, the fixed point is
\beq 
(u,v)=\left(\frac{\beta}{3}(-2\cosh\eta+1),0\right).
\eeq
It is easy to show that
$u$ is negative for this fixed point. The solution curves in the 
neighborhood of this fixed point will be unphysical.

\subsection{Case 2: $0<\Delta<1$} 

For this case, the solutions of equation (\ref{eq:y}) are
\beq
y_1=\cos \xi, 
\eeq
and 
\beq
y_{2,3}= -\cos\left(\frac{\pi}{3}\pm\xi\right), 
\eeq
where $\xi$ satisfies
\beq 
\cos 3\xi=\Delta=\frac{27\beta_1-2\beta^3}{2\beta^3}\label{eq:cos3xi}
\eeq
 and $0<\xi<\frac{\pi}{6}$. Therefore, by Equation (\ref{eq:u}), the  
three fixed points are 
\beq
(u_1,v_1)=\left(\frac{\beta}{3}(-2\cos\xi+1),0\right)
\eeq
\beq
(u_2,v_2)=\left(\frac{\beta}{3}(2\cos(\frac{\pi}{3}+\xi)+1),0\right),
\eeq
 and
\beq
(u_3,v_3)=\left(\frac{\beta}{3}(2\cos(\frac{\pi}{3}-\xi)+1),0\right).
\eeq

It can be shown that $u_1$ is negative, so 
the solution curves near the first fixed point, $(u_1,v_1)$, is unphysical.

We can also prove that the second fixed point, $(u_2,v_2)$, is 
an unstable saddle point and the third fixed point is a stable
center point (See Appendix B).



\subsection{Case 3: $-1<\Delta<0$} 

The solutions of equation (\ref{eq:y}) are
\beq
y_1=-\cos \phi,
\eeq
and 
\beq
y_{2,3}=\cos\left(\frac{\pi}{3}\pm\phi\right),
\eeq
where $\phi$ satisfies
\beq
\cos 3\phi=-\Delta=-\frac{27\beta_1-2\beta^3}{2\beta^3}\label{eq:cos3phi}
\eeq
 and $0<\phi<\frac{\pi}{6}$. 
Therefore, the fixed points are 
\beq
(u_1,v_1)=(\frac{\beta}{3}(2\cos\phi+1),0),
\eeq
\beq
(u_2,v_2)=(\frac{\beta}{3}(-2\cos(\frac{\pi}{3}+\phi)+1),0),
\eeq
 and
\beq
(u_3,v_3)=(\frac{\beta}{3}(-2\cos(\frac{\pi}{3}-\phi)+1),0).
\eeq

We can prove that the first fixed point, $(u_1,v_1)$, is 
an stable center point and the second fixed point, $(u_2,v_2)$ 
is an unstable
saddle point (See Appendix C).

It can be easily shown that $u_3$ is negative, so 
the solution curves near the third fixed point, $(u_3,v_3)$, is unphysical.

\section{The Results}

The phase plane analysis in the last section classifies the solution curves
by the relative strength parameter $\Delta$
(or parameters $\beta$ and $\beta_1$). 
 
There are therefore plenty of 
possible orbits during the scattering, especially for Case 2 and Case 3.
Because both Case 2 and 3 have one stable center and one unstable saddle
point in the $u>0$ region,   
we only need to use the results in Case 2 to demonstrate the
possible migrations of the planet in this section.


Fig.1 shows that a planet initially moving on a orbit with semi-major
axis $a=0.94$ can migrate to $a=1.57$ and the eccentricity increases from 
$e=0.33$ to $e=0.68$. 
(One can regard the unit of length to be AU in this paper.) 
The $u-v$ plane in Fig. 1(a) shows that there are two fixed points:
 one is an unstable saddle point and another is a stable center point. 
Fig.1(b)-(d)
are the orbital evolution on the $x-y$ plane and the
time variation for both the eccentricity and semi-major axis for the planet.
The solid line is for
the pre-scattering stage, the dotted line is for the scattering
stage and the dashed line is for the post-scattering stage.
The scattering comes in when $\theta= 2 \pi$ and leaves when $\theta= 3\pi$,
so the scattering acts for half of the orbital period. The dotted line in
Fig1(a) is the solution curve used for the scattering stage
in Fig.1(b)-(d). 


In order to simulate the scattering which takes place at 
larger semi-major axis, we choose the value
of $\beta$ to make the initial semi-major axis of the planet
to be around 24
and the initial eccentricity to be zero. 
Fig.2 shows that the planet initially moving on a orbit with semi-major
axis $a= 23.8$ can migrate to $a=24.8$ and the eccentricity increases 
from $e=0.0$ to $e=0.2$. 
Fig.2(b)-(d)
are the orbital evolution on the $x-y$ plane and the
time variation for both the eccentricity and semi-major axis for the planet.
The scattering begins when $\theta= 2 \pi$ and stops when $\theta= 3\pi$,
so the scattering acts for half of the orbital period. The dotted line in 
Fig.2(a) is the solution curve used for the scattering stage
in Fig.2 (b)-(d). The planet does not migrate as much as in Fig.1 because
the solution curve during the scattering stage is much closer to the
fixed point and the variation in $u$ (and therefore $r$) is much smaller.

To model a migration caused by a stronger scattering, we now assume
that there is velocity discontinuity when the planet enters
the scattering stage from the pre-scattering stage.
Fig.3 shows that a planet initially moving on a orbit exactly the same
as the initial orbit in Fig.2  can migrate 
to $a=69.0$ and the eccentricity increases from 
$e=0.0$ to $e=0.81$. 
Fig.3 (b)-(d)
are the orbital evolution on the $x-y$ plane and the
time variation for both the eccentricity and semi-major axis for the planet.

The scattering starts when $\theta= 2 \pi$ and stops when 
$\theta= 2.55 \pi$ so the scattering acts for about quarter 
of the orbital period. The dotted line in Fig3(a), the $u-v$ plane, 
is the solution curve used for the scattering stage in Fig.3(b)-(d).




We found that the eccentricity always increases and does not decay
to zero. 
In this model,
the only possibility to make the planet move on a 
orbit with zero eccentricity again is
if the scattering stage stops
precisely when the solution curve arrives at the fixed point of the 
post-scattering stage. This case can be neglect because 
the probability is too small. 

Because in the simulational model of Thommes, Duncan and Levison (1999)
the scattering takes place for many times before Neptune leaves the
Jupiter-Saturn region, one might want to generalize our model to
multiple scattering. 
One can choose small $\beta_1$ for each process and set $\beta_1=0$ 
between two scattering. All the results of phase plane analysis are still
the same and one only needs to pick up the correct solution curves to use
and connect them together. 

\section{Conclusions}

We have studied the dynamics of a planet under the influence of 
``swing scattering''. 
The governing equations can be 
transformed to an autonomous system and the standard phase plane analysis
was used to classify the parameter space and possible solutions.

When $\Delta >1$, i.e. $\beta_1>4\beta^3/27$, 
there is only one fixed point, where $u$ is negative. The solution curves
near this fixed point is therefore unphysical.
                               
When $0<\Delta<1$, i.e. 
$2\beta^3/27< \beta_1<4\beta^3/27 $, 
there are two fixed points where $u>0$. 
The two eigenvalues corresponding to $u_2$ 
are real with opposite sign, so $u_2$ is an unstable saddle point. 
The two eigenvalues corresponding to $u_3$ are pure complex 
with opposite sign, so $u_3$ is a stable center point. 

When $-1<\Delta<0$, i.e. $0<\beta_1<2\beta^3/27$, 
there are two fixed points where $u>0$. 
The two eigenvalues corresponding to $u_1$ are pure complex 
with opposite sign, so $u_1$ is a stable center point. The two eigenvalues
corresponding to $u_2$ are real with opposite sign, so $u_2$ is an
unstable saddle point.                                    

Thus,
we found that the solution topology is the same 
when 
$-1< \Delta < 1$ (with stable fixed point at physical region)
but becomes very different when $\Delta > 1$ 
(without stable fixed point at physical region)

Physically, this means when $\Delta > 1$, the scattering is so strong 
such that the planet will be forced to leave this system if the scattering
continues, no matter what position on the phase plane is when the scattering
happens. When $-1< \Delta < 1$, the scattering is weaker 
and the planet has chances to stay in this system even if the scattering
continues, depending on what position on the phase plane 
is when the scattering begins.

Therefore,  $\Delta > 1$ is an important instability criteria 
for ``swing scattering''.

These analytical results enable  us to study all possible orbits of 
outward planetary migration due to ``swing scattering''.
We found that the planet can migrate from an orbit with non-zero 
eccentricity to another eccentric orbit for both inner and outer 
part of the planetary system. In some situation, when 
the scattering is stronger, the planet (or small bodies
like Kuiper Belt Objects) can migrate to 100 AU. 
  
Our results reconfirm that the migration caused by scattering
always involves the eccentricity 
increasing during the orbital evolution. 
There is only one possibility that the orbit can be completely circular
again after the scattering stage. That is if the scattering stage stops
precisely when the solution curve arrives at the fixed point of the 
post-scattering stage. We can neglect this 
because the probability is extremely small. 

On the other hand, the orbit of the planet (Neptune) 
could be circularized
if it interact with the disc (Kuiper Belt) in the outer 
planetary system (Solar System) after the orbital scattered migration
as we see in the simulations of Thommes, Duncan \& Levison (1999).

If this was the process for the outer Solar System, i.e. 
Neptune was scattered to migrate outward but increased the 
eccentricity at first, then circularized by the Kuiper Belt afterward, 
the Kuiper Belt
has to be massive enough. This might indirectly supports the picture 
that the Kuiper Belt 
was at least two order more massive: 
an argument from the collision and formation history of the Kuiper 
Belt. If that is true, a large fraction of the Kuiper Belt Objects (KBOs)
would escape from the region of 30-50 AU and part of those non-escaped KBOs
would be captured by the Neptune into 3:2 resonance. 
The more complete physical picture
might be that the Neptune and the massive KBOs were coupled dynamically.
They influenced each other, i.e. KBOs circularized the Neptune's orbit and the 
Neptune scattered some of the KBOs,  until most KBOs left and the 
remained KBOs can then be treated as test particles under the 
influence of the outward migration of Neptune. This picture might be important 
for further study of the orbital evolution for both Neptune and KBOs.

\section*{Acknowledgment}
We are grateful to the referee's excellent suggestions. 

\section*{REFERENCES}
\begin{reference}

\reference Fernandez, J.A. \& Ip, W.-H. 1984, Icarus, 58, 109

\reference Goldstein H. 1980, Classical Mechanics, Addison-Wesley 
Publishing Company 

\reference Hahn, J.M. \& Malhotra, R. 1999, AJ, 117, 3041

\reference Jiang, I.-G. \& Ip, W.-H. 2001, A\&A, 367, 943

\reference Malhotra, R. 1995, AJ, 110, 420

\reference Neumark, S. 1965, Solution of Cubic and Quartic Equations,
Oxford, New York: Pergamon Press

\reference Thommes, E.W., Duncan, M.J. \& Levison, H.F. 1999,
Nature, 402, 635

\end{reference}

\clearpage

\appendix
\section{The Case when $\Delta=0$}

From (\ref{eq:3}), we know the eigenvalues satisfy
\beq
\left(\frac{\pa f_1}{\pa u}-\lambda\right)
\left(\frac{\pa f_2}{\pa v}-\lambda\right)-
\left(\frac{\pa f_1}{\pa v}\right)\left(\frac{\pa f_2}{\pa u}\right)=0.
\eeq

Since $f_1(u,v)=v$ and $f_2(u,v)=-u+\beta-\frac{\beta_1}{u^2}$, the above
equation becomes
\beq
\lambda^2-\left(-1+\frac{2\beta_1}{u^3}\right)=0.
\eeq
Thus,
\begin{eqnarray}
\lambda &=& \pm\left(-1+\frac{2\beta_1}{u^3}\right)^{1/2}, \nonumber \\
&=&\pm\left(\frac{-u^3+2\beta_1}{u^3}\right)^{1/2}.\label{eq:lambda}
\end{eqnarray} 

The solutions of Equation (\ref{eq:y}) are 
\beq
y_1=0,
\eeq 
and 
\beq
y_{2,3}=\pm\frac{\sqrt{3}}{2}.
\eeq
Therefore, from Equation (\ref{eq:u}), three fixed points are 
\beq
(u_1,v_1)=(\frac{\beta}{3},0),
\eeq
\beq
(u_2,v_2)=(\frac{\beta}{3}(1+\sqrt{3}),0),
\eeq
\beq
(u_3,v_3)=(\frac{\beta}{3}(1-\sqrt{3}),0).
\eeq

From Equation (\ref{eq:lambda}),
the eigenvalues corresponding to $(u_1,v_1)$ are $\pm\sqrt 3$ and
the eigenvalues corresponding to $(u_2,v_2)$ are
$\pm\sqrt {-1+\frac{4}{(1+\sqrt 3)^3}}$.  Thus, fixed point 
$(u_1,v_1)$ 
is an unstable saddle point, and $(u_2,v_2)$ is a stable center point.

 Since $u_3$ is negative, the solution curve in the neighborhood of  
$(u_3,v_3)$  is unphysical. 
                              
\section{The Eigenvalues for Case 2}

From Equation (\ref{eq:cos3xi}), we have
\beq
\beta_1=\frac{2}{27}\beta^3\cos3\xi+\frac{2}{27}\beta^3.
\eeq
 Let 
\begin{eqnarray*}
H_1(\xi,\beta)&\equiv &u_2-(2\beta_1)^{\frac{1}{3}}\\
&=&\frac{\beta}{3}(2\cos(\frac{\pi}{3}+\xi)+1)
-\frac{\beta}{3}(4\cos3\xi+4)^{\frac{1}{3}} \\
&=&\frac{\beta}{3}\left\{2\cos(\frac{\pi}{3}+\xi)+1-
(4\cos(3\xi)+4)^{\frac{1}{3}}\right\}.
\end{eqnarray*}
Thus,
\beq
\frac{\pa H_1}{\pa \xi}=\frac{\beta}{3}\left\{
-2\sin(\frac{\pi}{3}+\xi)+4\sin3\xi(4\cos3\xi+4)^{-\frac{2}{3}}\right\}.
\eeq

Numerically, we can easily show that 
\beq
-2\sin(\frac{\pi}{3}+\xi)+4\sin3\xi(4\cos3\xi+4)^{-\frac{2}{3}}<0, 
\eeq
for  $0<\xi<\frac{\pi}{6}$. Further, because $\beta>0$, we then have
$$\frac{\pa H_1(\xi,\beta)}{\pa\xi}<0.$$   
We also have $H_1(0,\beta)=0$, so it implies $H_1(\xi,\beta)<0$. 
Therefore, $u_2^3-2\beta_1<0$
for all $\beta>0$ and $0<\xi<\frac{\pi}{6}$.

Thus, from Equation (\ref{eq:lambda}), 
the two eigenvalues corresponding to $u_2$  
are real with opposite sign, so it is an unstable saddle point.

Since $0<\xi<\frac{\pi}{6}$ and 
$(\frac{\beta_1}{4})^{\frac{1}{3}}
<\frac{\beta}{3}<(\frac{\beta_1}{2})^{\frac{1}{3}},$
 we have 
$$2<2\cos(\frac{\pi}{3}-\xi)+1<\sqrt{3}+1.$$
Thus,
$$u_3=\frac{\beta}{3}\left(2\cos(\frac{\pi}{3}-\xi)+1\right)>\frac{2}{3}\beta
>2(\frac{\beta_1}{4})^{\frac{1}{3}}= (2\beta_1)^{\frac{1}{3}}.$$
Thus, $u_3^3-2\beta_1>0$.
 From Equation (\ref{eq:lambda}), the two eigenvalues corresponding to $u_3$ 
are pure complex with opposite sign, so it is a stable center point.

\section{The Eigenvalues for Case 3}

Since $0<\beta_1<\frac{2}{27}\beta^3$ and $0<\phi<\frac{\pi}{6}$ in 
this case, it implies 
\begin{eqnarray*}
u_1^3-2\beta_1&=&\frac{\beta^3}{27}(2\cos\phi+1)^3-2\beta_1\\
&>&\frac{\beta^3}{27}\{(2\cos\phi+1)^3-4\} \\
&>&\frac{\beta^3}{27}
\{(\sqrt{3}+1)^3-4\}\\
&>& 0
\end{eqnarray*}

Therefore, from Equation (\ref{eq:lambda}),
 the two eigenvalues corresponding to $u_1$ are pure complex 
with opposite sign, so it is a stable center point.   
 
From Equation (\ref{eq:cos3phi}), we have 
\beq
\beta_1=-\frac{2}{27}\beta^3\cos3\phi+\frac{2}{27}\beta^3.
\eeq
 Let
\begin{eqnarray*}
H_2(\phi,\beta)&\equiv&u_2^3-2\beta_1\\
&=&\left(\frac{\beta}{3}\right)^3
(-2\cos(\frac{\pi}{3}+\phi)+1)^3-2\beta_1 \\
&=&\frac{\beta^3}{27}\{(-2\cos(\frac{\pi}{3}+\phi)+1)^3
+4\cos(3\phi)-4\}.
\end{eqnarray*}

Follow the same approach as in Appendix B, we can show that
$H_2(\phi,\beta)<0$,
for all $\beta>0$ and $0<\phi<\frac{\pi}{6}$.
Therefore, we have $u_2^3-2\beta_1<0$.         
From Equation (\ref{eq:lambda}),
the two eigenvalues corresponding to $u_2$ are real with opposite sign, 
so it is an  unstable saddle point.

\clearpage

\begin{figure}[tbhp]
\epsfysize 7.0in \epsffile{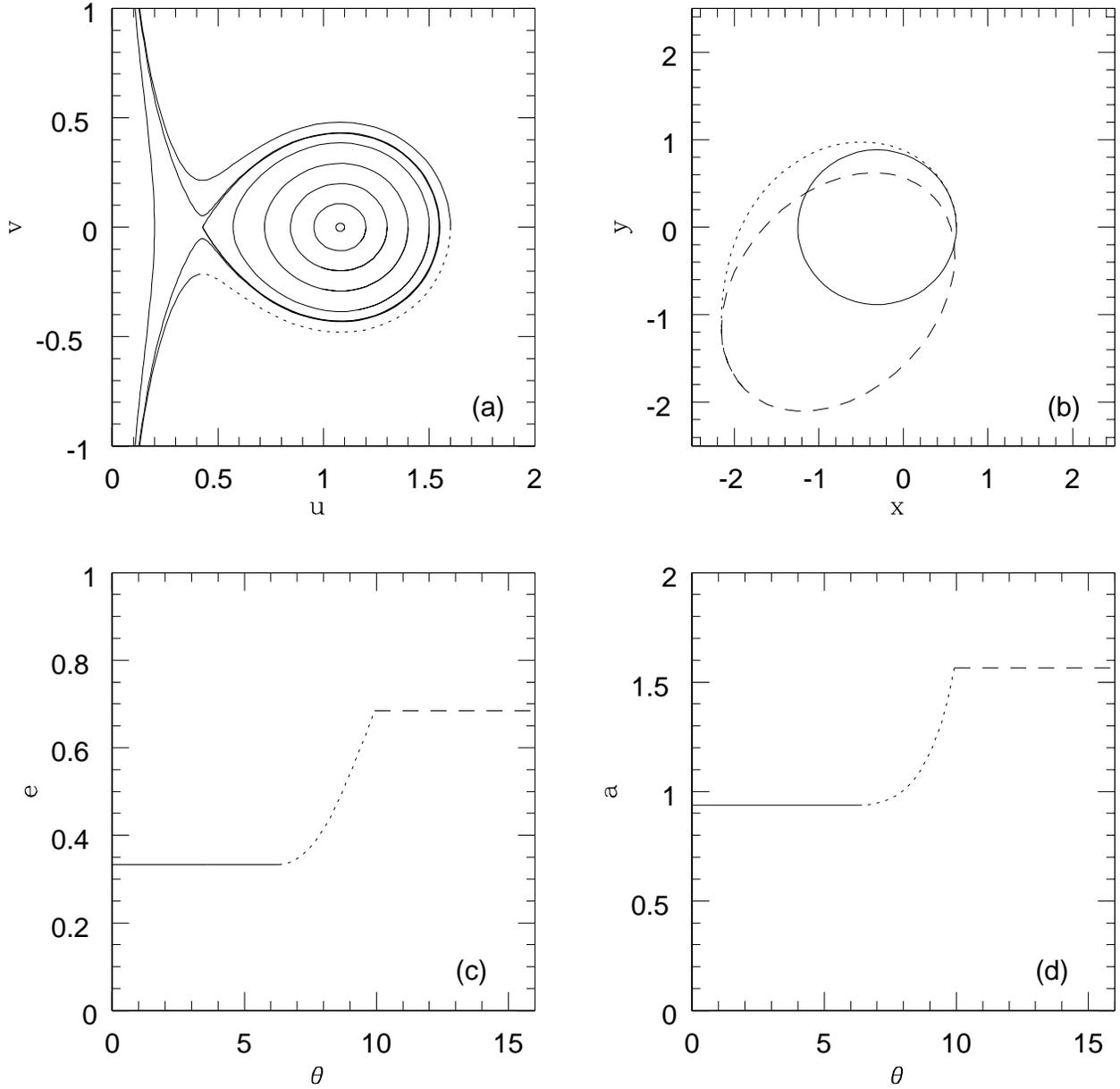}
\caption{
The Result when $\beta=1.2$, $\beta_1 =
2.2\beta^3/27$.
(a) The $u-v$ phase plane, where the dotted line is the solution
curve used during the scattering stage.
 (b) The orbit on the 
$x-y$ plane, (c) The eccentricity as function of $\theta$,
(d) The semi-major axis as function of $\theta$,
 where the solid line is for the pre-scattering 
stage, the dotted line
is for the scattering stage and
the dashed line is for the post-scattering stage.
}
\end{figure}

\begin{figure}[tbhp]
\epsfysize 7.0in \epsffile{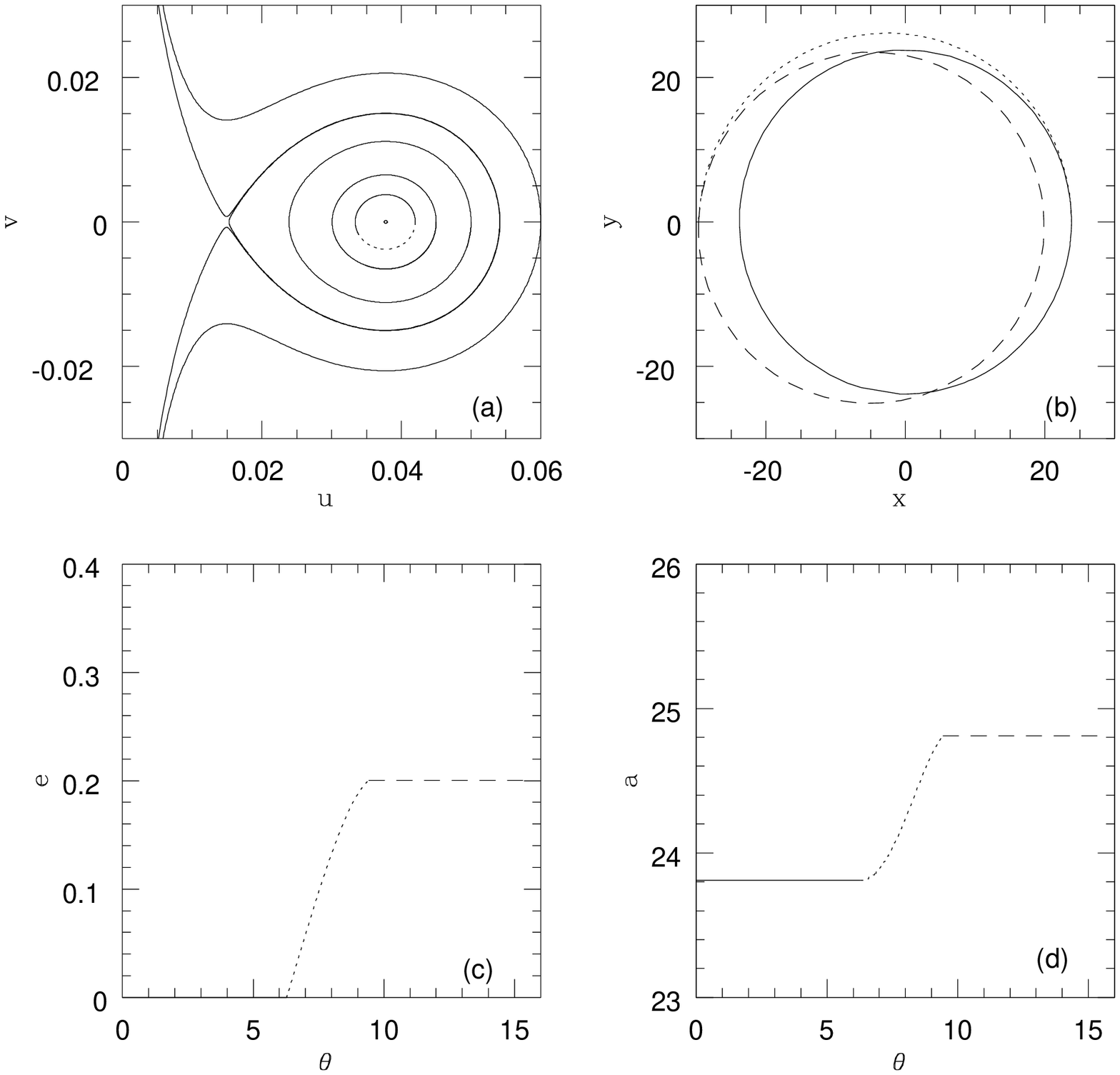}
\caption{
The Result when $\beta=0.04$, $\beta_1 =
2.2\beta^3/27$.
(a) The $u-v$ phase plane, where the dotted line is the solution
curve used during the scattering stage.
(b) The orbit on the 
$x-y$ plane, (c) The eccentricity as function of $\theta$,
(d) The semi-major axis as function of $\theta$,
 where the solid line is for the pre-scattering 
stage, the dotted line
is for the scattering stage and
the dashed line is for the post-scattering stage.
}
\end{figure}

\begin{figure}[tbhp]
\epsfysize 7.0in \epsffile{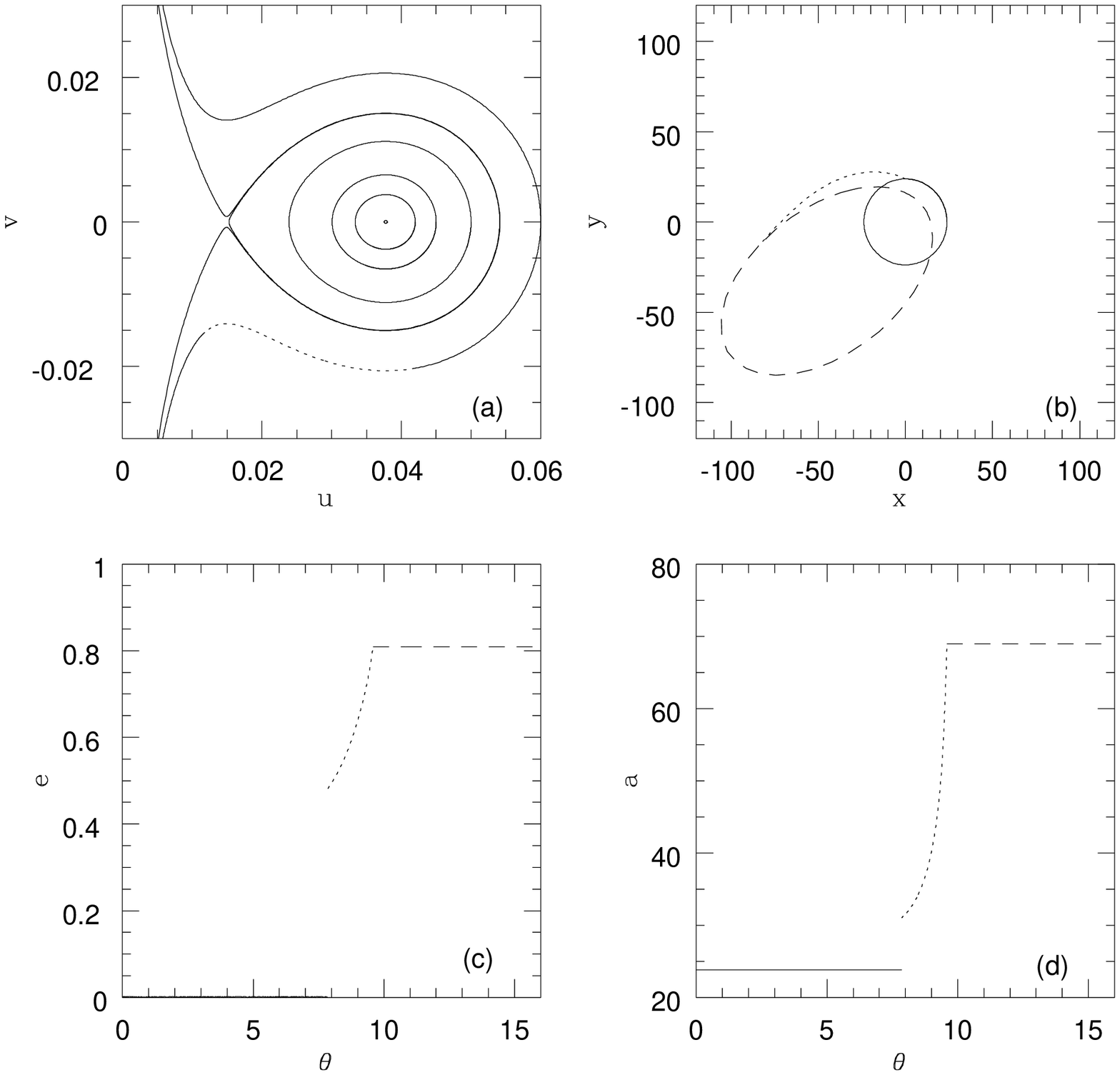}
\caption{
The Result when $\beta=0.04$, $\beta_1 =
2.2\beta^3/27$.
(a) The $u-v$ phase plane, where the dotted line is the solution
curve used during the scattering stage.
(b) The orbit on the 
$x-y$ plane, (c) The eccentricity as function of $\theta$,
(d) The semi-major axis as function of $\theta$,
 where the solid line is for the pre-scattering 
stage, the dotted line
is for the scattering stage and
the dashed line is for the post-scattering stage.
}
\end{figure}


\end{document}